\documentclass[12pt,epsf]{article}
\usepackage{epsf}
\usepackage{graphics}

\begin{document}
\title
{Restrictions on a geometrical language in gravity}
\author
{By Michael A. Ivanov \\
Chair of Physics, \\
Belarus State University of Informatics and Radioelectronics, \\
6 P. Brovka Street,  BY 220027, Minsk, Republic of Belarus.\\
E-mail: ivanovma@gw.bsuir.unibel.by.}
%\date{April 17, 2003}
\maketitle

\begin{abstract}
It was shown by the author (gr-qc/0207006) that screening the
background of super-strong interacting gravitons creates Newtonian
attraction if single gravitons are pairing and graviton pairs are
destructed by collisions with a body. In such the model, Newton's
constant is connected with Hubble's constant, for which the
estimate is obtained: $94.576 \  km \cdot s^{-1} \cdot Mpc^{-1}.$
It is necessary to assume an atomic structure of any body to have
the working model. Because of it, an existence of black holes
contradicts to the equivalence principle in a frame of the model.
For usual matter, the equivalence principle should be broken at
distances $\sim 10^{-11} \ m,$ if the model is true.
\end{abstract}

\section[1]{Introduction }
An alternative explanation of cosmological redshift \cite{1,2}
gives us a possibility to explain observed dimming of supernovae
Ia \cite{3} and the Pioneer 10 anomaly \cite{4} as additional
manifestations of the graviton background which is considered in a
flat space-time. It leads to a new cosmological model based on
this approach \cite{4a}. It was also shown that pressure of
correlated gravitonic pairs, which are destructed by collision
with a body, may create Newtonian attraction \cite{4b,4c}.
\par
In this paper, some important features of this approach are
described which give the restrictions on a geometrical language in
gravity. Gravity in this model may not be geometry at short
distances $\sim 10^{-11} \ m.$ At such the distances quantum
gravity cannot be described alone but only in some unified manner.
The geometrical description of gravity should be a good
idealization at big distances only by the condition of "atomic
structure" of matter. This condition cannot be accepted for black
holes which must interact with gravitons as aggregated objects.
The equivalence principle is roughly broken for black holes, if
this quantum mechanism of classical gravity is realized in the
nature.

\section[2]{Effects due to the graviton background}

In was shown in author's papers \cite{1,2} that a quantum
interaction of photons with the graviton background would lead to
redshifts of remote objects; the Hubble constant $H$ should be
equal in this model to:
\begin{equation}
H= {1 \over 2\pi} D \cdot \bar \epsilon \cdot (\sigma T^{4}),
\end{equation}
where $\bar \epsilon$ is an average graviton energy, $\sigma$ is
the Stephan-Boltzmann constant, $T$ is an effective temperature of
the graviton background assumed to be Planckian, and $D$ is some
new dimensional constant. To cause the whole redshift magnitude,
the interaction should be super-strong - it is necessary to have
$D \sim 10^{-27} m^{2}/eV^{2}.$ In the model, a photon energy $E$
should depend on a distance from a source $r$ as
\begin{equation}
                      E(r)=E_{0} \exp(-ar),
\end{equation}
where $E_{0}$ is an initial value of energy. It must be $a=H/c,$
where $c$ is the light velocity, to have the Hubble law for small
distances.
\par
Additional photon flux's average energy losses on a way $dr$ due
to rejection of a part of photons from a source-observer direction
should be proportional to $badr,$ where the factor $b$ is equal
to: $b=3/2 + 2/\pi = 2,137.$
\par
Such the relaxation together with the redshift will give, in a
case of flat no-expanding universe, the luminosity distance
$D_{L}$ \cite{3}, which is equal in our model to:
\begin{equation}
D_{L}=a^{-1} \ln(1+z)\cdot (1+z)^{(1+b)/2} \equiv
a^{-1}f_{1}(z;b).
\end{equation}
Comparison of the redshift model with supernova cosmology data
\cite{3} gives us a possibility to evaluate $H$ in our model (see
Fig. 1). Instead of prompt fitting to data, we can compare
$f_{1}(z;b)$ with one of the best fit of them. The function
$D_{L}(z; H_{0},\Omega_ {M},\Omega_{\Lambda}) \equiv
a^{-1}f_{2}(z; \Omega_{M},\Omega_{\Lambda})$ from \cite{3} (see
Eq.2 in \cite{3}) with $\Omega_{M}=-0.5$ and $\Omega_{\Lambda}=0,$
which is unphysical in the original work, gives such the fit. One
can see plots of the functions $f_{1}(z;b),$ $f_{1}(z;b)/1.09$ and
$f_{2}(z; \Omega_{M},\Omega_{\Lambda})$ on Fig. 1. The ratio
$H/H_{0}=1.09$ corresponds to the function $f_{1}(z;b)/1.09$ which
approximates $f_{2}(z; \Omega_{M},\Omega_{\Lambda})$ well enough
\cite{1,2}.
\begin{figure}[th]
\epsfxsize=12.1cm \centerline{\epsfbox{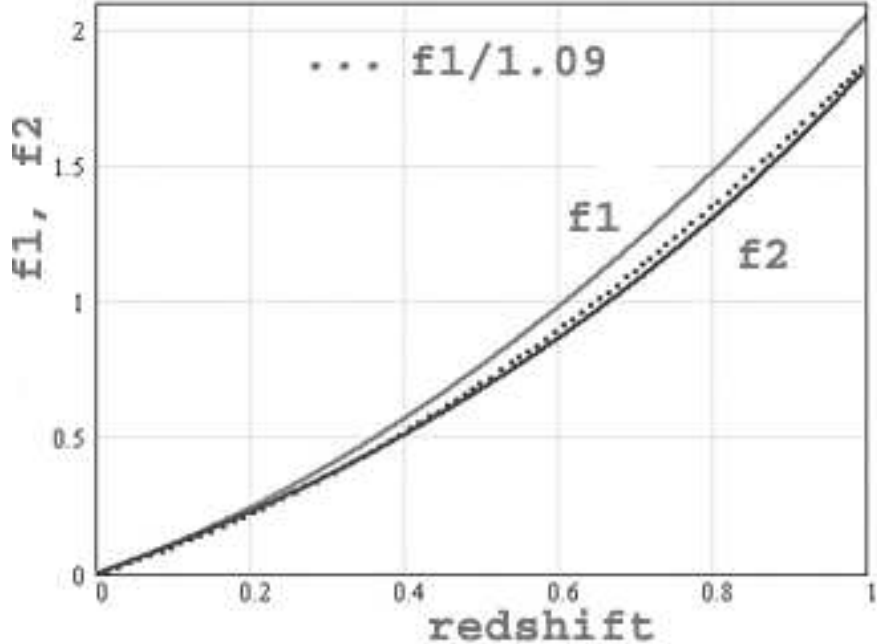}} \caption{
Comparison with supernova cosmology data. The functions
$f_{1}(z;b),$ $f_{1}(z;b)/1.09$ from this model, and $f_{2}(z;
\Omega_{M},\Omega_{\Lambda})$ giving one of the best fit to
supernova cosmology data \cite{3} are shown.}
\end{figure}
\par
But the very different functions $f_{1}(z;b)$ and $f_{2}(z;
\Omega_{M},\Omega_{\Lambda})$ may not be proportional to each
other on a big interval of redshifts. As one can see on Fig. 2, on
the interval $z \in (1,2)$ their plots are intersecting; after it,
the function $f_{2},$ values of which are connected only with
known supernova cosmology data for $z < 1,$ becomes much bigger
than $f_{1}.$

\begin{figure}[th]
\epsfxsize=12.98cm \centerline{\epsfbox{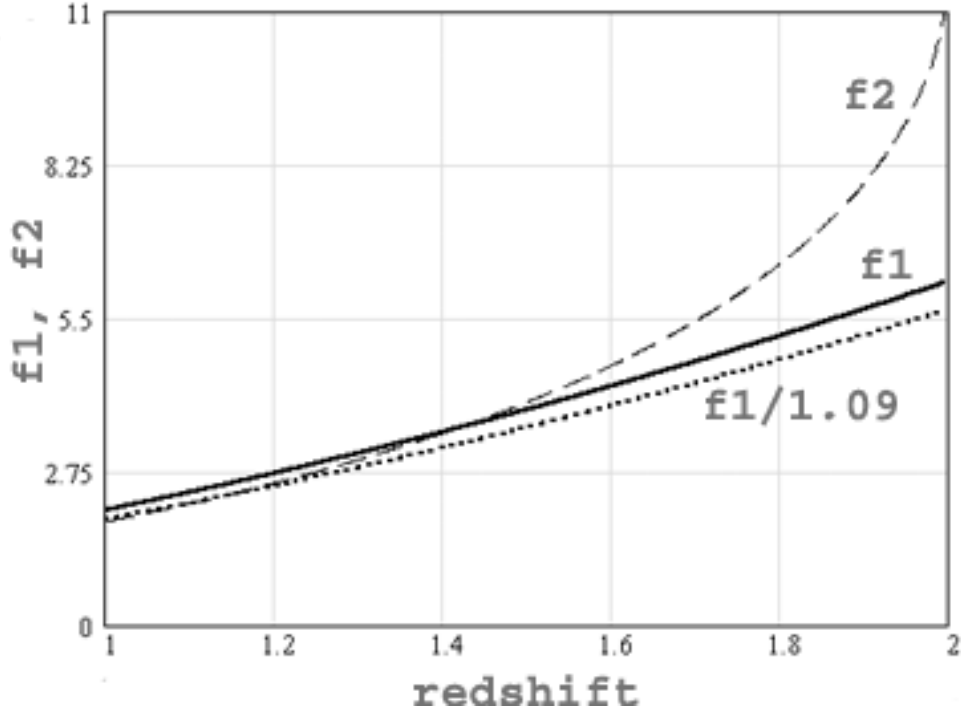}} \caption{The
same functions as on Fig. 1, but for bigger values of $z.$}
\end{figure}

It might mean, if one assumes that the present model is true, that
further investigations of supernovae Ia for bigger $z$ will lead
to some difficulties in interpretation of data in a frame of
cosmological models with expansion of the universe.
\par
The known conclusion about an accelerating expansion of the
universe \cite{3,33} is model-dependent. As one can see here, an
alternative explanation of dimming of supernovae Ia exists. This
explanation does not require an existence of any dark energy or
other exotic.
\par
Any massive objects, moving relative to the background, should
loss their energy too due to such a quantum interaction with
gravitons. It turns out \cite{1,2} that massive bodies must feel a
constant deceleration of the same order of magnitude as a small
additional acceleration of NASA cosmic probes \cite{4}. We get for
the body acceleration $w \equiv dv/dt$ by a non-zero velocity:
\begin{equation}
w = - ac^{2}(1-v^{2}/c^{2}).
\end{equation}
It is for small velocities: $w \simeq - Hc.$
\par
To ensure an attractive force bigger than a repulsive one due to a
presence of the graviton background, we need of graviton pairing
\cite{4}. In such the case, we get Newton's law in which Newton's
constant $G$ is equal to:
\begin{equation}
G = {2 \over 3} \cdot {D^{2} c(kT)^{6} \over {\pi^{3}\hbar^{3}}}
\cdot I_{2}
\end{equation}
where $I_{2}=2.3184 \cdot 10^{-6}.$
\par
We can establish in a frame of this model a connection between the
two fundamental constants $G$ and $H:$
\begin{equation}
H= (G  {45 \over 32 \pi^{5}}  {\sigma T^{4} I_{4}^{2} \over
{c^{3}I_{2}}})^{1/2},
\end{equation}
where $I_{4}=24.866.$ This connection gives us the following
estimate of Hubble's constant: $H= 3.026 \cdot 10^{-18}s^{-1},$ or
in the units which are more familiar for many of us: $H=94.576 \
km \cdot s^{-1} \cdot Mpc^{-1}$ (I would like to remember that
there is not any age of the universe in this model). This value of
$H$ is significantly larger than we see in the majority of present
astrophysical estimations \cite{3,12}, but it is well consistent
with some of them \cite{12a} and with the observed value of
anomalous acceleration of Pioneer 10 \cite{4} $w=(8.4 \pm
1.33)\cdot 10^{-10} \ m/s^{2}.$

\section[5]{Why and when gravity is geometry}
The described quantum mechanism of classical gravity \cite{4b,4c}
gives Newton's law with expression (5) for the constant $G$ and
the connection (6) for the constants $G$ and $H$ if the condition
of big distances is fulfilled:
\begin{equation}
\sigma (E,<\epsilon>) \ll 4 \pi r^{2},
\end{equation}
where $\sigma (E,<\epsilon>)= D \cdot E \cdot <\epsilon>$ is a
cross-section of interaction of a graviton with an average energy
$<\epsilon>$ with a body having an energy $E$ (there are two
different average energies in this model; for more details, see
\cite{4b}).

Newton's law is a very good approximation for big bodies of the
solar system. But assuming $r=1 \ AU$ and $E=m_{\odot}c^{2},$ we
obtain:
$${\sigma (E_{2},<\epsilon>) \over 4 \pi r^{2}} \sim 4
\cdot 10^{12}. $$ It means that in the case of interaction of
gravitons or graviton pairs with the Sun in the aggregate, the
considered quantum mechanism of classical gravity could not lead
to Newton's law as a good approximation. One must assume that
gravitons interact with "small particles" of matter - for example,
with atoms. If the Sun contains of $N$ atoms, then $\sigma
(E,<\epsilon>)=N \sigma (E_{a},<\epsilon>),$ where $E_{a}$ is an
average energy of one atom. For rough estimation we assume here
that $E_{a}=E_{p},$ where $E_{p}$ is a proton rest energy; then it
is $N \sim 10^{57},$ i.e. ${\sigma (E_{a},<\epsilon>)/ 4 \pi
r^{2}} \sim 10^{-45} \ll 1.$
\par
This necessity of "atomic structure" of matter for working the
described quantum mechanism is natural relative to usual bodies.
But would one expect that black holes have a similar structure? If
any radiation cannot be emitted with a black hole, a black hole
should interact with gravitons as an aggregated object. For bodies
without an atomic structure, the allowances, which are
proportional to $D^{3}/ r^{4}$ and are caused by decreasing a
gravitonic flux due to the screening effect, will have a factor
$m_{1}^{2}m_{2}$ or $m_{1}m_{2}^{2}.$ These allowances break the
equivalence principle for such the bodies.
\par
For bodies with an atomic structure, a force of interaction is
added up from small forces of interaction of their "atoms": $$ F
\sim N_{1}N_{2}m_{a}^{2}/r^{2}=m_{1}m_{2}/r^{2},$$ where $N_{1}$
and $N_{2}$ are numbers of atoms for bodies $1$ and $2$. The
allowances to full forces due to the screening effect will be
proportional to the quantity: $N_{1}N_{2}m_{a}^{3}/r^{4},$ which
can be expressed via the full masses of bodies as
$m_{1}^{2}m_{2}/r^{4}N_{1}$ or $m_{1}m_{2}^{2}/r^{4}N_{2}.$ By big
numbers $N_{1}$ and $N_{2}$ the allowances will be small. Let us
denote as $\Delta F$ an allowance  to the force $F.$ Then we get
by $E=E_{\odot}, \ r=1 \ AU$ that the ratio ${\Delta F \over F}
\sim 10^{-46}.$
\par
One can replace $E_{p}$ with a rest energy of very big atom - the
geometrical approach will left a very good language to describe
the solar system. We see that for bodies with an atomic structure
the considered mechanism leads to very small deviations from the
equivalence principle, if the condition (7) is fulfilled for
microparticles, which prompt interact with gravitons.
\par
For small distances we shall have:
\begin{equation}
\sigma (E,<\epsilon>) \sim 4 \pi r^{2}.
\end{equation}
It takes place by $E_{a}=E_{p}, \ <\epsilon> \sim 10^{-3} \ eV$
for $r \sim 10^{-11} \ m.$ This quantity is many order larger than
the Planck length. The equivalence principle should be broken at
such distances.
\par
Under the condition (8), big digressions from Newton's law will be
caused with two factors: 1) a screening portion of a running flux
of gravitons is not small and it should be taken into account by
computation of the repulsive force; 2) a value of this portion
cannot be defined by the expression $\sigma (E,<\epsilon>) / 4 \pi
r^{2}.$
\par
One might expect that a screening portion may tend to a fixing
value at super-short distances. But, of course, at such distances
the interaction will be super-strong and our naive approach would
be not valid.

\section[6]{Conclusion}
Observations of last years give us strong evidences for
supermassive black holes in galactic nuclei. Of course, a central
dark mass in galactic nucleus may not be a black hole; it is most
likely to the one by its properties from all known objects. We
must remember that we know only that these objects are
supermassive and compact - and we, further, suggest that they are
black holes. It was supposed by the author \cite{4a} that such
black holes may be collectors of virtual massive gravitons and
"germs" of galaxies in a frame of this approach. But the analysis
of \cite{4b,4c} shows that black holes should interact with
gravitons as aggregated objects and, therefore, their existence
must break the equivalence principle. We have a dilemma here: to
accept their existence and breaking the equivalence principle or
to find other candidates on a role of observable supermassive and
compact objects in the approach. For example, virtual massive
gravitons might be collected in some kind of Bose condensate.
\par
A main conclusion of my mentioned works is that, if a redshift has
the non-Dopplerian nature, an interaction of a graviton with any
particle should be super-strong. This circumstance would lead to a
new manner to construct a quantum description of gravity and to
unify known interactions.

\end{document}